\documentclass[a4paper,fleqn,usenatbib]{mnras}

\usepackage{graphicx}   
\usepackage{amsmath}    
\usepackage{amssymb}    
\usepackage{subfig}
\usepackage{color}

\usepackage[T1]{fontenc}
\usepackage{ae,aecompl}

\title[The radial distribution of BSS in GCs]{Investigating the Blue Straggler Stars radial distribution in globular clusters with Monte Carlo simulations}
\author[Sollima et al.]{A. Sollima$^{1}$\thanks{E-mail:
antonio.sollima@inaf.it}, 
F. R. Ferraro$^{2}$\\
$^{1}$ INAF Osservatorio di Astrofisica e Scienza dello spazio di Bologna, via Gobetti 93/3, 40129 Bologna, 
Italy\\
$^{2}$ Dipartimento di Fisica e Astronomia, Universit\'a di Bologna, via Gobetti 93/2, 40129 Bologna, Italy\\
}

\date{Accepted 2018 November 27. Received 2018 November 26; in original form 2018 June 29}

\pubyear{2017}

\begin{document}
\label{firstpage}
\pagerange{\pageref{firstpage}--\pageref{lastpage}}
\maketitle


\begin{abstract}
We investigate the evolution of the radial distribution of Blue Straggler 
Stars through a set of Monte Carlo simulations of star clusters under a variety 
of initial conditions.
We used a novel technique based on the "artificial oversampling" of the 
distribution function of the Blue Stragglers and control population to tear down the effect of 
statistical fluctuations affecting the determination of the relative distribution 
of these stellar populations. We find that a bimodal distribution, qualitatively 
similar but much less pronounced than those observed in many globular clusters, 
can naturally emerge as a result of the 
progressive migration of Blue Stragglers in the energy domain. The behaviour of the 
parameter $A^{+}$, proposed as a "dynamical age" indicator,
 has been also investigated. This parameter shows
 a relatively homogeneous and well defined trend with the fraction of the 
 elapsed core-collapse timescale up to the core collapse phase, 
 while after this stage its evolution depends on initial conditions.
\end{abstract}

\begin{keywords}
methods: numerical -- methods: statistical -- blue stragglers -- stars: kinematics and dynamics --
globular clusters: general 
\end{keywords}

\section{Introduction}
\label{intro_sec}

The dynamical evolution of globular clusters (GCs) is a complex topic determined by the
interplay of several physical processes like stellar evolution, stellar
collisions, tidal interactions, etc.
Although the origin of GCs is still controversial
\citep{1968ApJ...154..891P,1985ApJ...298...18F,2008MNRAS.391..825D}, 
the major milestones of 
their dynamical evolution have been largely studied through several kinds of 
dynamical simulations
\citep{1971Ap&SS..14..151H,1989ApJ...342..814C,1991ApJ...370..567G,2003MNRAS.343..781G,2007MNRAS.374..344T,2007ApJ...658.1047F}.

On the other hand, from the observational point of view, we can only derive 
information on the present-day structure and kinematics of a given GC observing a snapshot of its 
entire evolution. For this reason, in the past years an important effort has
been put in the search for an indicator of the stage of dynamical evolution of a
GC acting as a "clock".

As the main process at work during GC evolution is the two-body relaxation, all
the indicators proposed so far are based on the effect left by this process on
the main GC observables. In particular, interactions lead to a tendency toward kinetic 
energy equipartition with the most massive stars sinking in less energetic
(inner) orbits with respect to low-mass stars.
As indicators of mass-segregation, the comparison of the mass
function slopes estimated at different
radii \citep{2017MNRAS.464.1977W} and the variation of the velocity dispersion as a
function of mass \citep{2018MNRAS.475L..96B} have been proposed.

In a series of papers the Bologna 
University group proposed that a special class of stars, the so-called Blue 
Straggler Stars (BSS) could be used to trace the internal dynamical evolution 
of clusters (see \citealt{1999ApJ...522..983F,2006ApJ...638..433F,2009Natur.462.1028F,2017ApJ...839...64R,2018ApJ...860...36F}). 
This peculiar population of stars has been observed in the colour-magnitude
diagram of all GCs at magnitudes
brighter than the Main Sequence turn-off, mimicking the presence of a population
of young massive stars. The main hypothesis for the origin of these stars imply
that they are the result of mass-transfer occurring in primordial
binaries \citep{1964MNRAS.128..147M} or in stellar collisions \citep{1976ApL....17...87H}.
Aside from their formation process, BSS are thought to be 3-4 times more massive 
than the average mass ($<m>=0.3-0.4 M_{\odot}$) of stars populating old stellar 
systems such as Galactic GCs and for this reason have been suggested to be suitable 
{\it gravitational probes} to explore the degree of mass segregation in these stellar systems. 
In particular, in order to study the BSS radial distribution 
\citet{1993AJ....106.2324F} compared the fraction of BSS with respect to a reference
population (usually Horizontal Branch or Red Giant Branch stars; $N_{BSS}/N_{con}$) and normalized to the
fraction of cluster integrated light sampled in the same bin ($R_{BSS}$). 

The radial distribution of BSS
has been found multi-modal with at least three different morphologies: 
{\it (i) Family I - flat distribution}: BSS are distributed following the cluster 
light with $R_{BSS}=1$ at any distance from the cluster center \citep{2006ApJ...638..433F,2008ApJ...681..311D}; 
{\it (ii) Family II - bimodal distribution}: with a central overdensity of BSS 
followed by a dip at intermediate radii and an increase at large distances 
\citep{1997A&A...320..757F,2004ApJ...603..127F}; 
{\it (iii) Family III - uni-modal distribution}: with a central peak and a 
monotonic decreasing at larger distance \citep{2012ApJ...748...91C}. 
The physical interpretation of such variety of morphologies has been provided by
\citet{2012Natur.492..393F} \citep[see also][]{2006MNRAS.373..361M} who proposed that the different 
observed morphologies are due to the radial dependence of the dynamical 
friction timescale producing a progressive migration of BSS toward the GC 
center and depriving regions at larger distance with increasing time.
In this scenario, the location of the minimum ($R_{min}$) in the $R_{BSS}$ 
radial distribution turns out an indicator of 
the completion of the BSS sedimentation process.

However, this indicator has two major drawbacks: {\it i)} it requires to bin the
radial extent of data with a consequent degree of arbitrariety, and {\it ii)} 
even the most massive GCs host only a few hundreds of BSS \citep{2006ApJ...638..433F},
thus it suffers of a large statistical noise.

Moreover, while the observational evidence of the BSS bimodal distribution is 
apparent in many GCs, its presence and prominence is less apparent in simulations.
The first attempt in detecting such a feature in an N-body simulation has been 
made by \citet{2015ApJ...799...44M}, who confirmed the general picture 
suggested by \citet{2012Natur.492..393F}: the formation of a sharp 
central peak, the development of a small dip ($(N_{BSS}/N_{RGB})_{min}\sim0.9(N_{BSS}/N_{RGB})_{tot}$) 
in the radial 
distribution and the evolution toward a monotonic distribution at advanced 
stages. However the detection of the bimodal distribution turned out to be 
quite noisy and the migration of the minimum toward large distance from the 
cluster center was difficult to be followed. 
Similar  difficulties have been found by \citet{2017MNRAS.471.2537H} who 
analysed several Monte Carlo simulations of GCs under different initial and environmental condition.
They found that, while the
BSS always segregate into the central regions of the GC, the
increasing trend at large distances has been found as a unstable feature
which also depends on the adopted binning. 

In order to provide a more easily observable indicator of the BSS 
sedimentation process
 \citet{2016ApJ...833L..29L} defined the parameter $A^{+}$ as the area between 
 the normalised radial cumulative distribution of the BSS and a reference 
 population. They found a significant correlation between $A^{+}$ and the 
 position of the minimum $R_{min}$ in the analysed GCs and a nice relation 
 with their cluster central 
relaxation time. \citet{2018ApJ...860...36F} extended this study to 48 GGC 
confirming the correlation between $A^{+}$ and the number of elapsed central 
relaxation times, thus suggesting the parameter $A^{+}$ as a powerful 
dynamical age indicator.
\citet{2016ApJ...833..252A} 
tested the
validity of this indicator in a set of N-body simulations run in idealized
conditions (without binaries and stellar evolution and assuming BSS as single 
massive
test particles) and found a good correlation between the $A^{+}$ parameter and
the fraction of elapsed core-collapse time ($t/t_{cc}$).

Note that the simulations by \citet{2017MNRAS.471.2537H}, \citet{2015ApJ...799...44M} and 
\citet{2016ApJ...833..252A}, draw their conclusions using a realistic
sample of only $\sim$100 BSS per simulation. Indeed, in a real GC BSS constitute 
only a tiny fraction of the GC mass, so that any improper oversampling of the
BSS population in a simulation would bias its predictions.

In this paper we investigate the radial distribution of BSS adopting a novel 
technique to overcome to the problem of the low statistics. In Sect. 2 we
discuss on the concept of "dynamical age" and provide our working definitions.
In Sect. 3 the technique of artificial distribution function oversampling and 
its application to our Monte Carlo code is introduced. Sect. 4 is devoted to 
the description of the performed simulations. The resulting radial distributions 
of BSS is presented in
Sect. 5. In Section 6 we discuss the effectiveness of the $R_{min}$ and $A^{+}$ 
parameters in measuring cluster dynamical evolution.
 We summarize and discuss our conclusions in Sect. 7.

\section{Considerations on the "dynamical clock"} 
\label{cons_sec} 

The search for  parameters tracing the "dynamical age"  of a stellar system is triggered by the following question:
{\it "how advanced is the dynamical evolution of a stellar system?"}.
The above question, although crucial, could be misleading, since different indicators 
can be advocated to trace different dynamical processes occurring
over different timescales affecting different regions of the cluster and all of 
them modifying the structural and kinematical parameters.

In particular, for a pressure supported stellar system with an age
comparable to or longer than its relaxation time, an initial phase of 
expansion \citep[driven by the potential energy losses related to the evolution of massive stars in the first $10^{8}$ yr;][]{1990ApJ...351..121C} is followed by a long period in which
the large number of interactions between stars lead massive stars to release part 
of their kinetic energy to less massive ones, moving on inner orbits. As a
result of this process, the mass function varies as a function of the
distance from the center and, in any given region of a GC, 
massive stars are kinematically colder than low-mass ones \citep{1969ApJ...158L.139S}.
At the same time, the large velocity acquired by low-mass stars make them more 
prone to evaporation with a consequent flattening of the global MF. Such a
process can be accelerated by the interaction with the Galactic tidal field
whose strength depends on the cluster orbit \citep{2003MNRAS.340..227B}. The interplay between two-body
relaxation and mass-loss leads to an instability of the core which progressively
loses kinetic energy and shrinks by more than an order of magnitude
\citep[core-contraction and -collapse;][]{1970MNRAS.150...93L}. The possible presence of binaries
or heavy remnants (like neutron stars and black holes; hereafter NS and BH,
respectively) reduces the degree of mass segregation and can delay (or even halt
and reverse) core-collapse since these stars quickly
sink into the cluster core and interact with other stars through three- and
four-body encounters in which part of the binding energy of binaries is
transferred into kinetic energy \citep{1975MNRAS.173..729H}. These systems, because of their large mass have 
low orbital energy and are therefore more efficiently retained with respect to
single low-mass stars.

The efficiency and the timescale of the above processes depend in a complex way on
the initial values of many parameters (i.e. mass-function, binary fraction 
and characteristics, cluster 
size and structure, orbit, etc.) which could have been different among GCs at 
their birth.

Moreover, processes like kinetic energy equipartition and core-collapse, while 
both being the outcome of the dynamical evolution, have efficiencies which depends
on different parameters being therefore not univocally correlated 
i.e. it is possible to find GCs close to core-collapse whose stars have
exchanged a lower fraction of their kinetic energy than those in a GC which is
still far from this stage.
So, although the general trend is toward high
degrees of mass segregation, concentrated cores, large remnant 
fractions, none of the above observables univocally describe all the aspects of
dynamical evolution. 

For example, the problem of the dynamical cluster evolution can be put 
as follows:
 {\it "the amount of changes in the structural and kinematical
properties of a GC from their initial conditions"}. This definition, while
intuitive, is practically unaffordable. Indeed, at present, there is not a clear
picture of the first stages of the GC life, where initial conditions are set. 
The proposed scenarios of GC formation involve a multi-phase gas collapse
forming multiple populations of stars characterized by different chemistry, structure and
kinematics \citep{2008MNRAS.391..825D}, the
interaction of proto-GCs in gas-rich molecular clouds \citep{2017MNRAS.467.1857B} and the possible
presence of a primordial dark matter halo \citep{1968ApJ...154..891P}. Besides this, it is 
unlikely that GCs form with universal initial 
conditions. Among the various parameters, many works suggested variations of the
initial mass function \citep{2010MNRAS.406.2000M}, initial size and concentration
\citep{2010MNRAS.401.1832B},
binary fraction \citep{2008MNRAS.388..307S}, and degree of primordial mass segregation \citep{2007AJ....134.1368C,2007ApJ...655L..45M},
whose existence is under debate. 
It is therefore not obvious that a single (or a
combination of) present-day observable can describe the amount of evolution 
of the same parameters in different GCs regardless of their unknown initial 
conditions.

A second possible definition is linked to {\it "the amount of energy exchanged
by stars through their mutual interactions"}. In the pioneering work by \citet{1969ApJ...158L.139S} this quantity has been expressed as a function of the timescale over 
which a "typical"
cluster star lose memory of its initial velocity. Following this definition, it is possible to
define the so-called "two-body relaxation time" as 
$$t_{rel}\sim0.35 \frac{\sigma^{3}}{G^{2}\langle m \rangle \rho~ln\Lambda}$$ 
where $\rho$ is the mass density, $\sigma$ is the 3D 
velocity dispersion, $\langle m \rangle$ is the mean stellar mass and
$ln \Lambda$ is the Coulomb logarithm. Note that, according to the above formula, the relaxation
time varies across the cluster being smaller in the center than in the
outskirts. To define an average distance-independent timescale, \citet{1969ApJ...158L.139S,1987degc.book.....S}
defined also the "half-mass relaxation time" defined as the relaxation time
calculated using the density calculated within the half-mass radius ($r_{h}$) and the
average cluster velocity dispersion
\begin{equation}
t_{rh}=\frac{0.138 M^{1/2}r_{h}^{3/2}}{G^{1/2} ln \Lambda}
\label{eq_trh}
\end{equation}
Note that, because of its arbitrary
definition, this last timescale provides only an order of magnitude of the
typical timescale of energy exchange within a cluster, neglecting any effect of 
structural differences (like e.g. concentration) and assuming an
isolated evolution driven only by two-body relaxation. Despite this
approximation the initial half-mass relaxation time is often used as reference in
theoretical works. However, the mass, the half-mass radius and the mean
mass change during the evolution, so that also this timescale varies with time.
This leads to two problems: {\it i)} from an observational point of view, it is
possible to determine only the present-day half-mass relaxation time which can
be different from the initial one, and {\it ii)} the actual timescale over which
stars have exchanged a significant fraction of their kinetic energy depends on
the whole evolution of the half-mass relaxation time and not on its
value at a given epoch.

A third possibility is to identify a milestone in the typical evolution of a GC
and define its "dynamical age" as {\it "the position of the cluster along the
path toward this reference epoch"}. Also this definition has many drawbacks.
Indeed, the natural reference epoch occurring during the evolution of a GC is
the core-collapse. However, while core-collapse is easily identifiable in
simulations composed only by single-mass particles, the presence of a primordial population of
binaries makes its identification more difficult. Indeed, while in the absence of
primordial binaries there is a clear transition between the phase of 
core-contraction and that where the first binaries (created through tidal 
capture) start to release energy, this transition does not occur when primordial 
binaries smoothly release part of their energy during the core-contraction
phase \citep{2007MNRAS.374..344T}. Moreover, the timescale over which core-collapse occurs
($t_{cc}$)
is proportional to the central relaxation time \citep[$t_{cc}\sim 300 t_{rel,0}$;][]{1995PASJ...47..561T}. So, concentrated GCs experience core-collapse even when
the majority of their stars (orbiting preferentially at large distances) have
still not significantly interacted with each others. This means that $t_{cc}$ and $t_{rh}$ are 
time-scales which are not univocally correlated. For the above reasons in the following analysis, when comparing different
simulations, we will express the time as a function of both the 
half-mass relaxation time measured at the epoch of the considered snapshot 
($t_{rh}(t)$), and the core-collapse time. We also introduce an additional adimensional 
parameter 
$$S(t)=\int_{0}^{t} \frac{dt'}{t_{rh}(t')}$$
which express the number of times a star has lost memory of its past motion. Thus, the three quantities $t_{cc}$, $t_{rh}$ and S represent time-scales
on which different dynamical phenomena occurs.

\section{Method}
\label{met_sec}

The simulations used in this paper were run using a modification of the Monte 
Carlo code presented in \citet{2014MNRAS.443.3513S}.
The general description of the orbit-averaged Monte Carlo method has been
originally provided by \citet{1971Ap&SS..14..151H} and later developed by many authors 
\citep[see e.g.][]{1982AcA....32...63S,1998MNRAS.298.1239G,2000ApJ...540..969J}.
Briefly, the cluster is considered as a sample of $N$ superstars characterized by
their mass, energy and angular momentum per unit mass which generate a spherical
symmetric potential. At each time-step the following steps are performed: 
\begin{itemize}
\item{a statistical realization of the cluster
is performed by placing superstars at random positions along their orbits 
according to the time spent at that distance from the cluster center;}
\item{each superstar is assumed to interact with its nearest neighbor producing
a perturbation on their energies and angular momenta;}
\item{the potential is evaluated according to the masses and positions of the
superstars.}
\end{itemize}
The above steps are repeated until the end of the simulation. 
The above scheme is efficient and versatile allowing to easily include many 
levels of complexity like the presence of a
mass spectrum \citep{2001MNRAS.324..218G,2001ApJ...550..691J}, the effects of stellar evolution \citep{2010ApJ...719..915C}, three- and four-body 
interactions \citep{2007ApJ...658.1047F} and interaction with a tidal field \citep{2014MNRAS.443.3513S}.
The effectiveness of the above technique in reproducing the effect of
two-body relaxation in a wide range of initial conditions and environments has 
been widely tested against N-body simulations \citep[see
e.g.][]{1975IAUS...69..133H,2001MNRAS.324..218G,2001ApJ...550..691J}.
A detailed
description of the above steps is provided in \citet{2014MNRAS.443.3513S} and will not be repeated here. Below we describe the modification to
their code made to increase the statistics of BSS without biasing the result of
simulations.

Each simulation is composed by three distinct samples: {\it i)} a {\it main 
sample}
constituted by 29000 stars with different masses, fraction of binaries and remnants, {\it ii)} a
{\it BSS sample}, constituted by 10000 binaries whose evolution naturally drives toward BSS
formation, and {\it iii)} a {\it control sample}, constituted by 10000 test particles with
the mass of a typical Red Giant Branch (RGB) star in a GC ($M=0.83~M_{\odot}$). 
To define the {\it BSS sample}, we considered the formation
scenario proposed by \citet{1964MNRAS.128..147M} in which BSS form as a result of the
Roche-lobe overflow occurring during the expansion of the primary component of a
close binary after its hydrogen exhaustion. For this purpose, we simulated a 
large population of binaries (following the prescription described in Sect. \ref{sim_sec}) 
and extracted only those systems satisfying the following conditions: {\it i)} a 
primary star with mass larger than the turn-off mass of a old ($t=12$ Gyr) 
metal-poor ($[Fe/H]=-1.3$) $\alpha$-enhanced isochrone \citep[from the][database]{2008ApJS..178...89D},
{\it ii)} a primary star radius at the RGB tip larger than the volume-averaged 
Roche lobe radius, following the criterion of \citet{1988ApJ...334..688L}, {\it iii)} a
total lifetime (primary + BSS) longer than 12 Gyr. The above criteria have been
verified at each timestep to exclude binaries which, after a series of
interactions, modified their characteristics and would not produce BSS.
Particles in the {\it main sample} evolve independently of the other samples
following the canonical steps described above. At each timestep, the particles
of the {\it BSS} and {\it control} samples are ranked in distance and associated
to the closest {\it main sample} particle. The perturbation in energy and
angular momentum are calculated and applied only to the stars in the  {\it BSS} and 
{\it control} samples\footnote{It is possible that in the same time-step two or 
more {\it BSS} and/or {\it control} particles interact with the same
{\it main sample} particle. This however does not produce any inconsistency in the
evolution of the energy and angular momentum distributions because {\it main
sample} particles do not feel the presence of the other samples and they provide only
the statistical probability of interaction with a given cluster star.}. 
In this way, particles in the {\it BSS} and {\it control} 
samples feel the effect of interactions with the cluster population but their
presence does not affect the canonical evolution of the {\it main} sample.
The possible occurrence of three- and four-body is also
accounted for according to the adopted prescriptions for this kind of interactions
(see below). Being subject to the same kind of interactions occurring 
with the same probability as their corresponding population in the {\it main
sample}, the {\it BSS} and {\it control} samples share the same
distribution in phase-space of their parent stars, while their sizes are much
larger. This novel technique, hereafter referred as "artificial
distribution function oversampling", allows to model the evolution of a number
of {\it BSS} and {\it control} sample particles which exceeds by orders of
magnitude those of real BSS and RGB stars, without affecting the conservation of
energy and angular momentum of the entire simulation\footnote{Note that each {\it BSS} and {\it control} sample particle can interact 
only with a particle of the {\it main} sample, while {\it BSS-BSS}, {\it control-control} and {\it BSS-control} interactions are forbidden. 
In this way, each particle of the {\it BSS} and {\it control} samples is an independent tracer of the distribution function of its population.}.
As a consequence, the
distribution function of {\it BSS} and {\it control} sample particles, and their
corresponding radial distributions, can be sampled with thousands of objects,
thus reducing statistical fluctuations. Of course, the relative fraction of
{\it BSS} and {\it control} sample particle do not reflect the corresponding fraction of
these objects in the {\it main sample}. Hence, only their phase-space distribution 
will be used to study the BSS radial 
segregation and to derive the "dynamical age" indicators proposed in the literature, 
which are all based on the radial distribution of BSS independently on the 
size of the BSS and reference population samples.
As a sanity check on the validity of the "artificial oversampling" technique, we 
compared the number fraction and energy distribution of particles in the 
{\it control sample} and those in the {\it main sample} with a mass in the range
$0.78<M/M_{\odot}<0.83$ at each timestep of all the simulations performed here.
These quantities indeed trace the evaporation rate and the efficiency of two-body 
relaxation, respectively, and the same trend must be found in the oversampled ({\it control}) and 
{\it main} samples in the same mass range.
The above defined number counts ratio remains constant during the simulations within the uncertainties,
although in the last snapshots the Poisson noise due to the low number of
particles in the {\it main sample} makes this ratio quite noisy and small
deviations ($<20\%$) can occur because of the slightly lower mean mass of
particles in the {\it main sample}. 
Moreover, we also performed a Kolmogorov-Smirnov test on both the cumulative radial
and energy distribution of {\it control sample} and {\it main sample} particles in the above
defined mass range in selected snapshots of each simulations. Again, we always find a
probability $>95\%$ that the particles are drawn from the same distribution in
both the radial and the energy domain. 
Unfortunately, the same test cannot be
performed on {\it BSS sample} particles because only a handful of BSS are
present in the {\it main sample} thus hampering the significance of such a test.

Additional modifications have been made on the original code described in
\citet{2014MNRAS.443.3513S} to account for the presence of a mass
spectrum and for the formation and
interactions involving binaries. The only modification needed in the presence of a spread of 
masses is in the definition of the Coulomb logarithm (defined as the
logarithm of the ratio between the maximum impact parameter and the $90^{\circ}$
deflection radius). This quantity enters in
both the definition of the initial half-mass relaxation time (see eq.
\ref{eq_trh}) and in the estimation of the effective deflection angle in each
interaction \citep[$\beta$; see][]{1971Ap&SS..14..151H} 
$$\beta=2~arcsin \left(\sqrt{(m_{1}+m_{2})^{2} n~ln\Lambda}\right)$$
where $m_{1}$ and $m_{2}$ are the masses of the interacting stars and $n$ is the
local number density. Unfortunately, there is not a precise definition of
the maximum impact parameter when determining the exact value of $\Lambda$ in a
general set of conditions, and various approaches have been used in the
literature \citep[see][for an extensive discussion on this
issue]{1989MNRAS.239..549W,1993ApJ...410..543W}.
At each timestep we determined $\Lambda$ using eq. 34 of \citet{1975IAUS...69..133H}, which 
has been converted into a discrete form   
$$\Lambda=0.11 \exp\left(\frac{N \displaystyle\sum_{i=0}^{N/2-1} m_{2i+1} m_{2i+2}^{2} ln \left(\frac{2 \langle m
\rangle}{m_{2i+1}+m_{2i+2}}\right)}{2 \displaystyle\sum_{i=0}^{N/2-1} m_{2i+1}
\displaystyle\sum_{i=0}^{N/2-1} m_{2i+2}^{2}}\right)$$
where $\langle m \rangle$ is the mean particle mass of the simulation.
We adopted the normalization factor 0.11 to converge to the value predicted by 
\citet{1994MNRAS.268..257G} in the case of an equal-mass particles simulation.
For reference, the above formula gives a value of $\Lambda\sim0.02$ when a 
\citet{2002Sci...295...82K} IMF is considered, in agreement with what found by 
\citet{1996MNRAS.279.1037G}.

To account for the binary formation and three/four-body interaction we defined the probability of
inelastic encounter between single-single, binary-single and binary-binary stars as
$$P=w~n~\pi r_{m}^{2} \left( 1+\frac{2 G (m_{1}+m_{2})}{w^{2} r_{m}}\right)
\Delta t$$
with
\[ r_{m}=
\begin{cases}
b_{crit} & for~single-single\\
3.5~a& for~binary-single\\
2~(a_{1}+a_{2})& for~binary-binary\\
\end{cases}
\]
where $w$ is the relative velocity of the objects, $a(_{1,2})$ is
the semi-major axis of the binary star(s) and $b_{crit}$ is the critical impact
parameter for to tidal capture \citep{1999A&A...347..123K}.

For each pair of neighbors, the appropriate probability has been calculated 
and a random number uniformly distributed
between 0 and 1 is extracted. If the random number is smaller than the
associated probability the close interaction is integrated uning a symplectic
integrator with adaptive time-step \citep{1990PhLA..150..262Y} which ensures a conservation
of energy at $\Delta E/E<10^{-9}$. For this purpose, the initial
conditions of the interaction are set by randomly rotating the binary rotation 
plane(s) and placing the interacting objects at a distance such that the
potential energy felt by the two objects is 20 times smaller than the binary
binding energy and with an impact parameter 
$$b=r_{m}\sqrt{\eta}$$ 
where $\eta$ is a random number uniformly distributed between 0 and 1. The particles move toward their 
center of mass with a relative velocity calculated from their relative 
velocity at infinity. The interaction is followed
until the total energy of the system exceeds zero. The products of the
collision are then placed with their final velocities in the cluster reference
frame. In the case of tidal capture, we assumed that the newly formed binary has
a semi-major axis $a=2~b$ and zero eccentricity, as expected as a result of
energy dissipation in very-close binaries \citep{1986ApJ...310..176L}.

\section{Description of simulations}
\label{sim_sec}

\begin{table*}
 \centering
 \label{tab:table1}
  \caption{Initial conditions of our simulations.}
  \begin{tabular}{@{}lccccccccr@{}}
  \hline
 Name & $W_{0}$ & $r_{c}(0)$ & IMF & $f_{b}$ & $\mu_{NS,BH}$ & $R_{G}$ & $t_{rh}(0)$ & $t_{cc}$ & colour\\
      &         & pc      &     &  \%     & \%            & kpc & Gyr & Gyr &\\ 
 \hline
 w5rg5  & 5      & 3       & Kroupa (2002) & 5 & 0 & 5 & 1.35 & 4.97 & black\\
 w5rg10 & 5      & 3       & Kroupa (2002) & 5 & 0 & 10 & 1.35 & 5.47 & red\\
 w3rg5  & 3      & 4.76    & Kroupa (2002) & 5 & 0 & 5 & 1.35 & 4.45 & green\\
 w5rg5nobin  & 5 & 3       & Kroupa (2002) & 0 & 0 & 5 & 1.35 & 5.20 & cyan\\
 w5rg5bh & 5     & 3       & Kroupa (2002) & 5 & 5.2 & 5 & 1.35 & 1.37 & blue\\
 w5rg5mf & 5     & 3       & power-law $\alpha=-1$ & 5 & 0 & 5 & 0.69 & 3.57 & magenta\\
\hline
\end{tabular}
\end{table*}

\begin{figure*}
 \includegraphics[width=\textwidth]{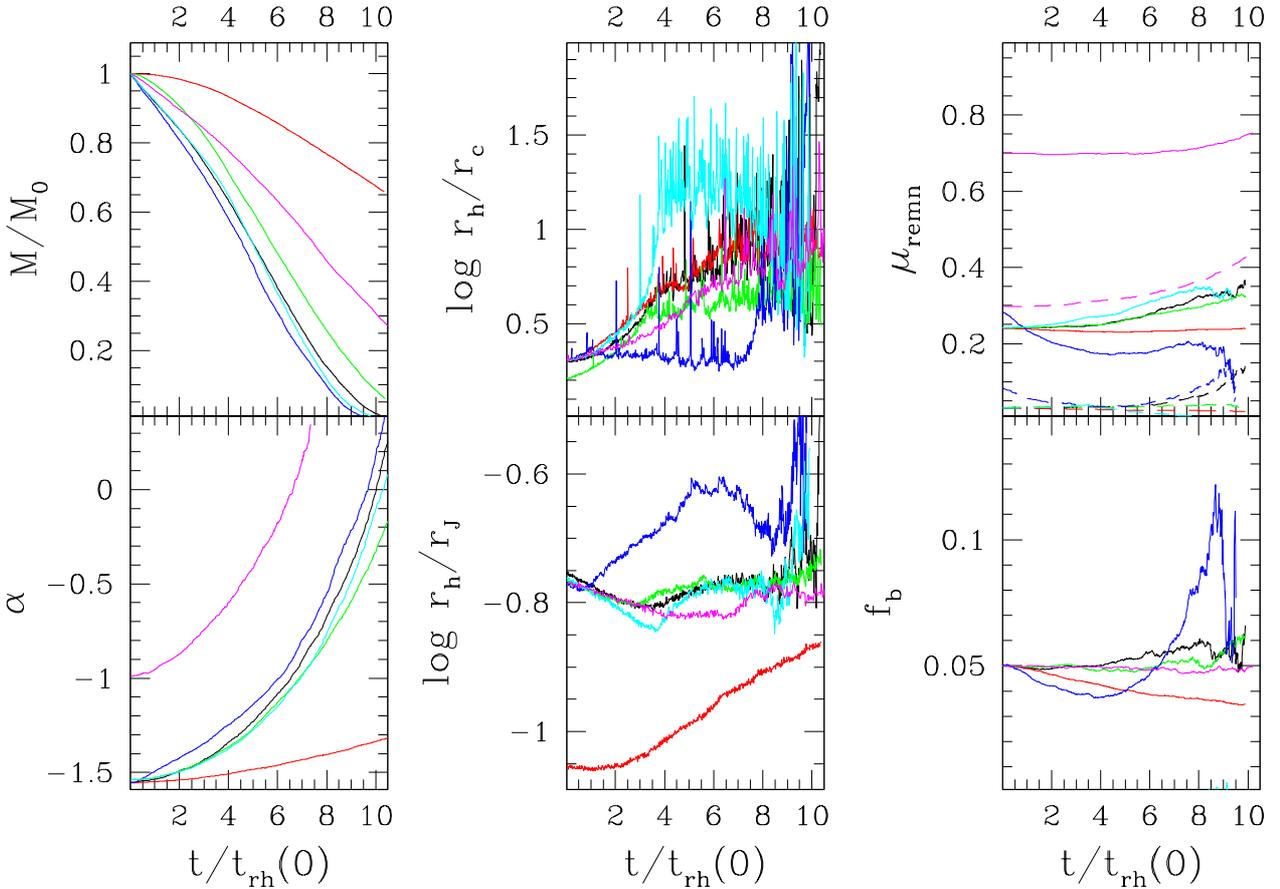}
 \caption{Evolution of the enclosed mass (top left panels), mass function slope (bottom left
 panels), core concentration ($log (r_{c}/r_{h})$; top central panels), 
 Roche-lobe filling factor ($log (r_{h}/r_{J})$; bottom central panels), fraction of remnants 
 (top right panels; the total and massive ($M>0.83~M_{\odot}$) remnant fractions are marked by solid and dashed lines, 
 respectively) and fraction of binaries (bottom right panels).
 The behaviour of simulation w5rg5 (black lines), w5rg10 (red), w3rg5 (green),
 w5rg5bh (blue) w5rg5nobin (cyan) and w5rg5mf (magenta) are shown.}
\label{all}
\end{figure*}

\begin{figure*}
 \includegraphics[width=\textwidth]{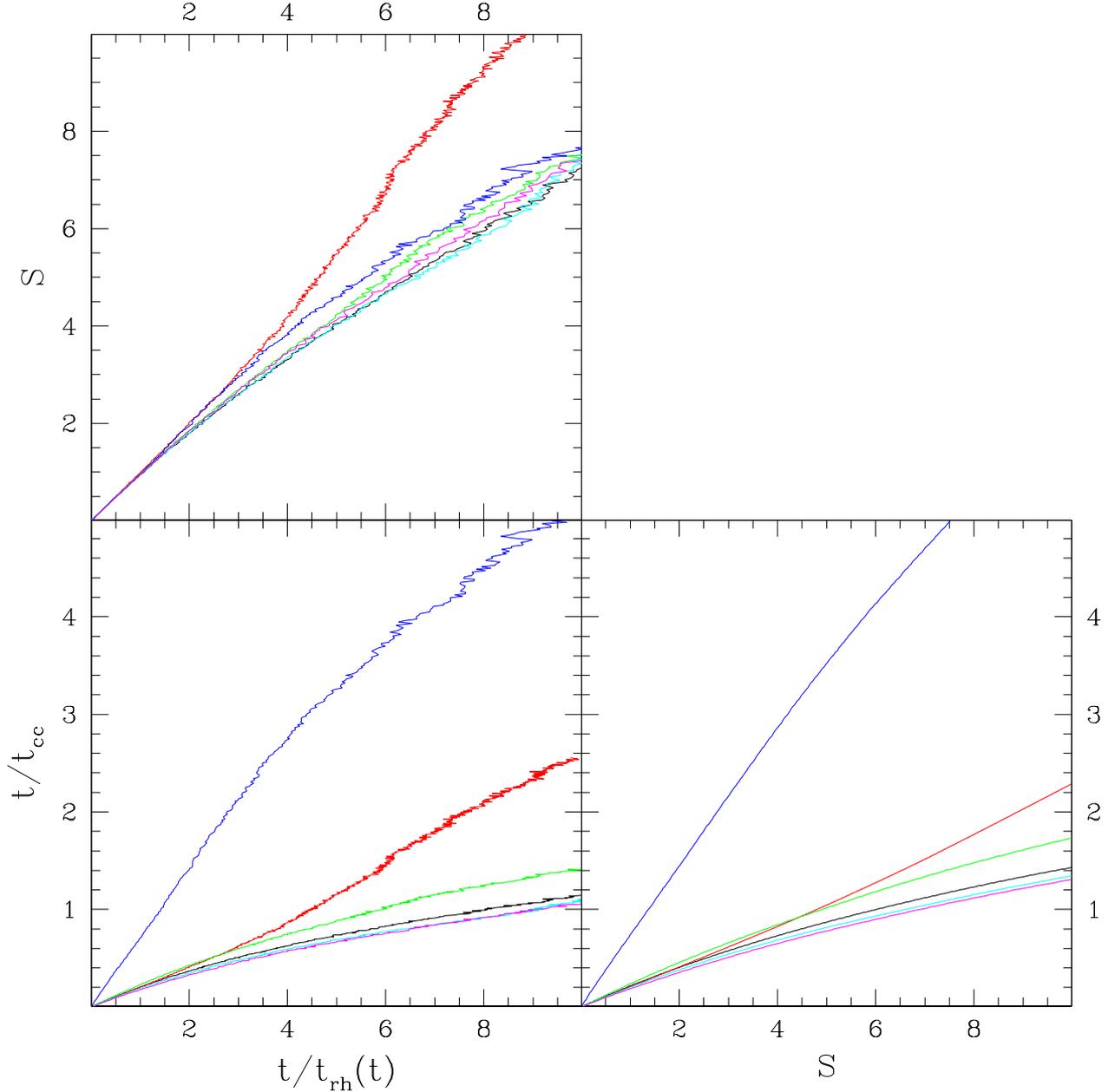}
 \caption{Comparison of the three dynamical timescales $t/t_{rh}(t)$, $S$
 and $t/t_{cc}$.
The adopted colour code is the same of Fig. \ref{all}.}
\label{time}
\end{figure*}

\begin{figure*}
 \includegraphics[width=\textwidth]{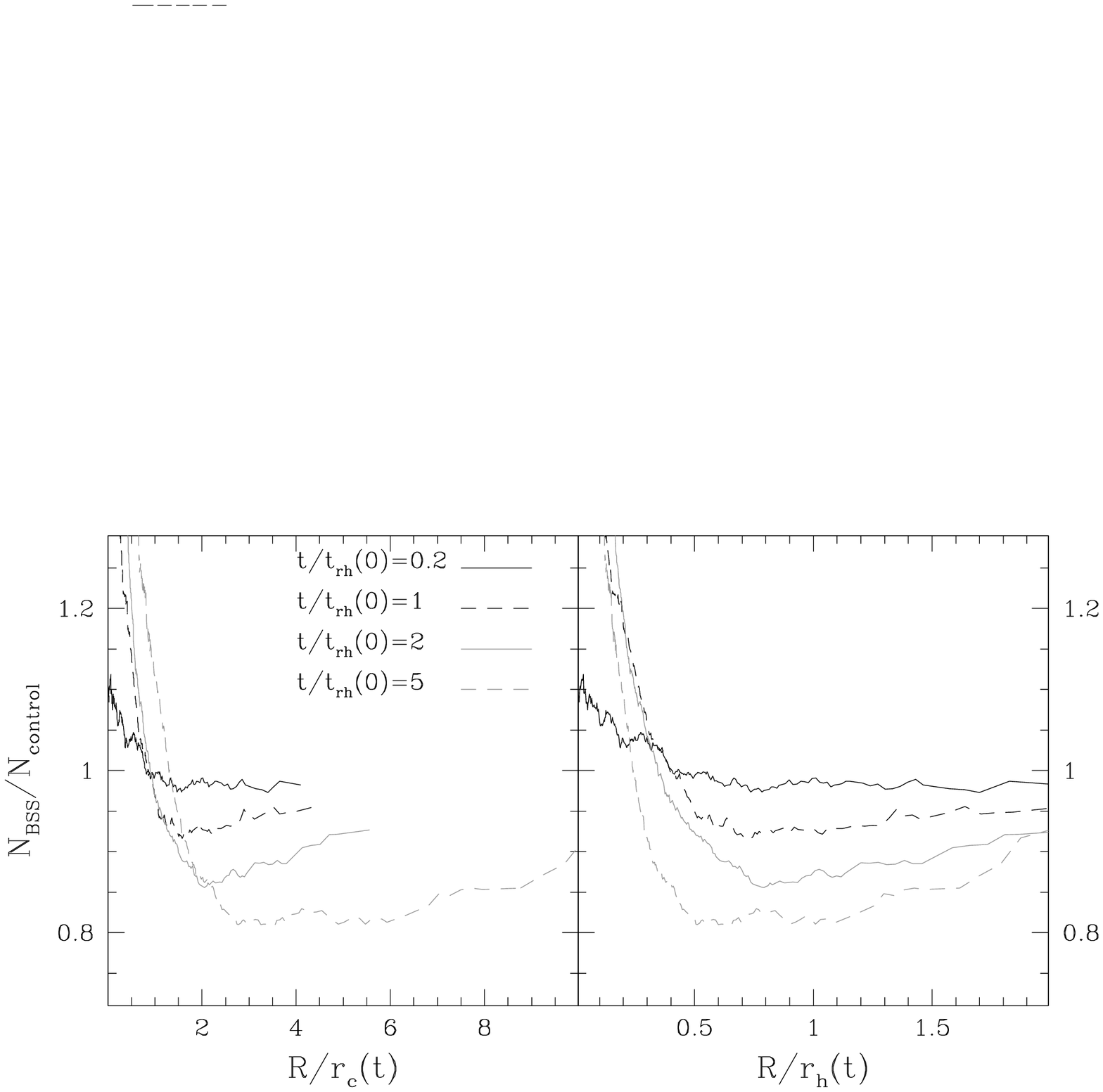}
 \caption{Normalized ratio of {\it BSS} and {\it control} sample particles in
 simulation w5rg5. The ratio is plotted both as a function of the core (left
 panel) and the half-mass (right panel) radius. Different lines indicates different stages of evolution. 
 The typical uncertainties in this plot are $<0.01$}
\label{rap}
\end{figure*}

\begin{figure*}
 \includegraphics[width=\textwidth]{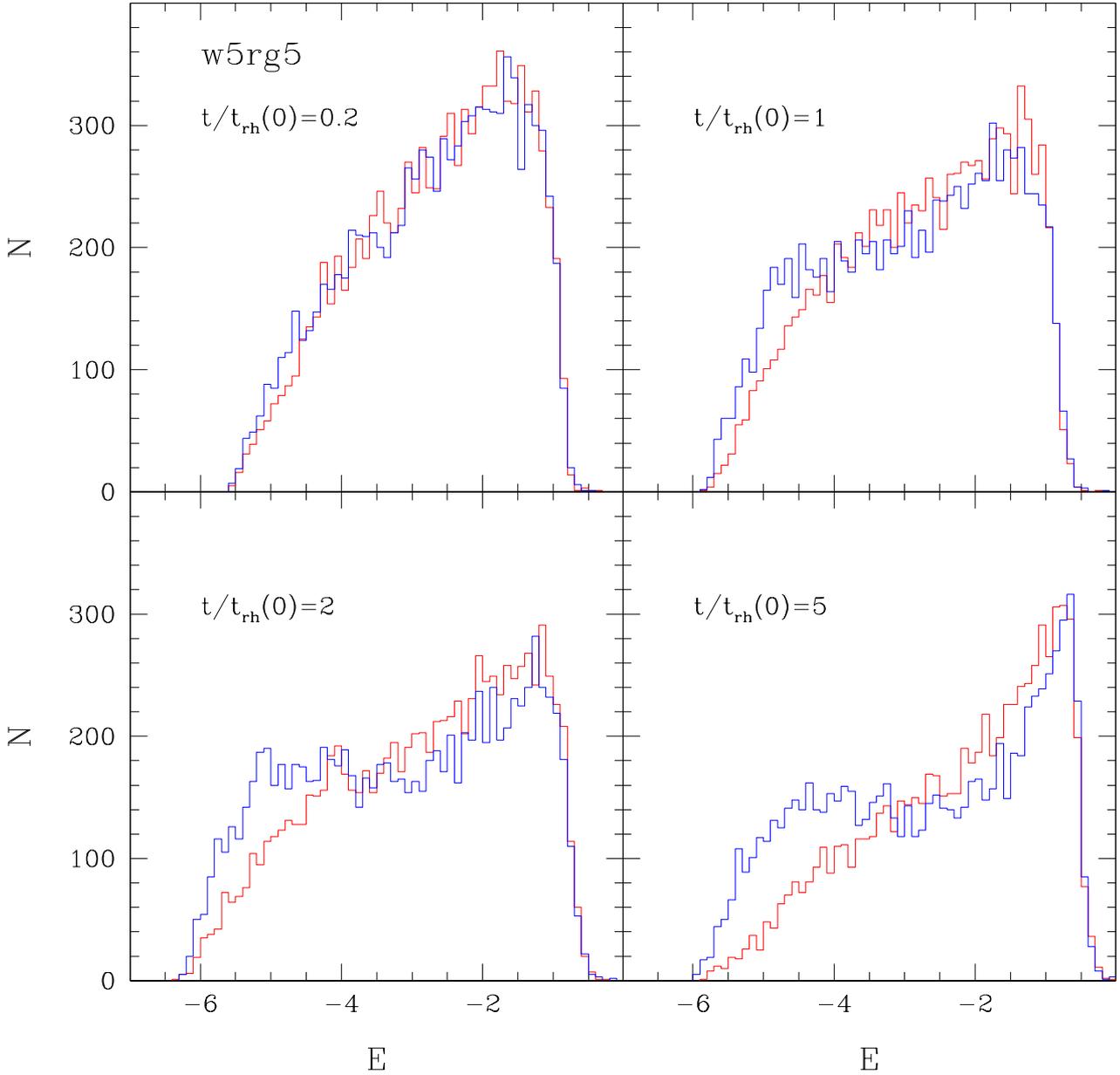}
 \caption{Distribution of orbital energies of {\it BSS} (blue histograms) and 
 {\it control} (red histograms) sample
 particles in different snapshots of the w5rg5 simulation.}
\label{ene}
\end{figure*}

\begin{figure}
 \includegraphics[width=8.6cm]{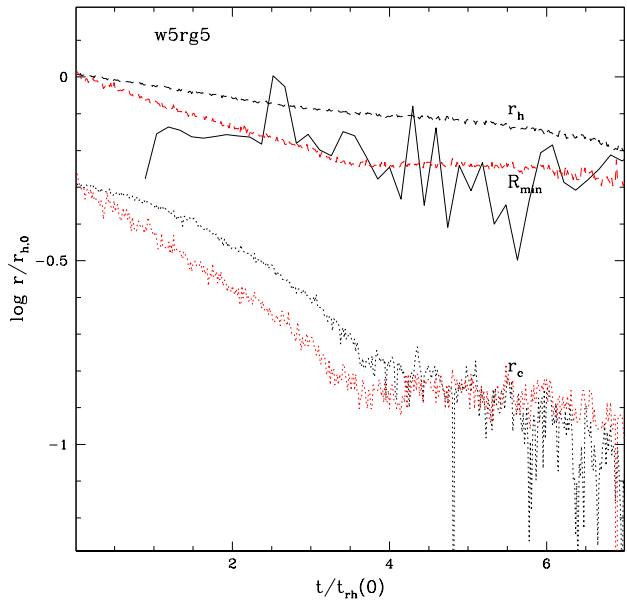}
 \caption{Evolution of the core, half-mass radius and radius of the minimum of the BSS distribution in simulation
 w5rg5. Black lines indicate the $r_{c}$ defined as in \citet{1985ApJ...298...80C} (dotted lines) 
 and the actual value of $r_{h}$ (dashed lines), while red lines are the same radii estimated through the 
 comparison with \citet{1966AJ.....71...64K} models.}
\label{rad}
\end{figure}

\begin{figure*}
 \includegraphics[width=\textwidth]{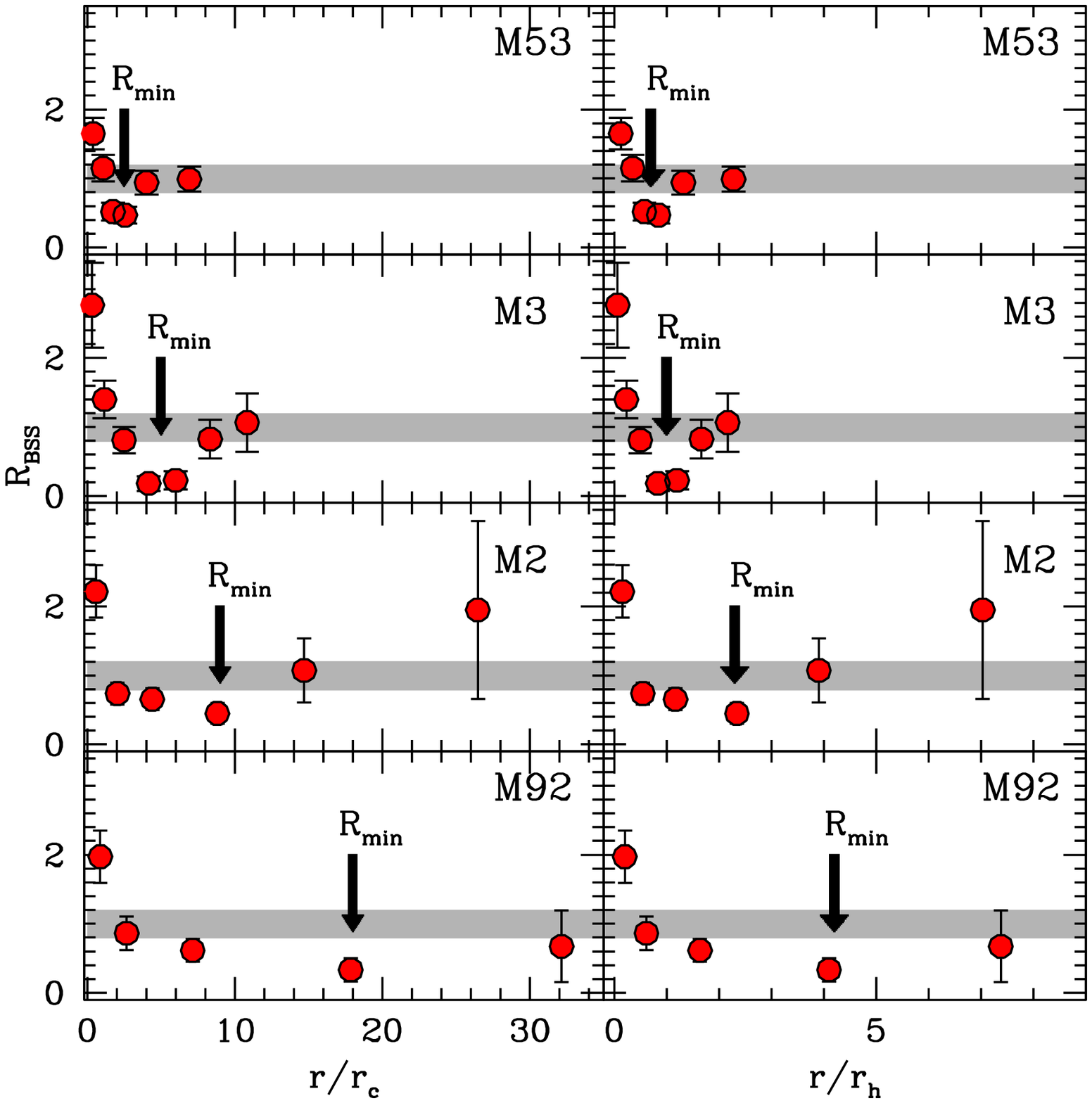}
 \caption{The normalised BSS radial distribution (red dots) for four clusters in the sample presented by \citet{2012Natur.492..393F} 
are plotted as function of the distance from the cluster center normalised to the core radius (left panels) and the half-mass radius (right panels).
the grey strips schematically show the distribution of the reference populations. The thick arrows show the approximate location of the minimum of the distribution.}
\label{rmin}
\end{figure*}

In principle, to properly simulate the outcome of the evolution of a GC at the
present day, simulations should start with a wide mass spectrum, evolve
for a Hubble time including the effects of stellar evolution, and then only their
last snapshot should be analysed. However, this would require a huge number of
simulations to sample different stages of dynamical evolution with a
corresponding unaffordable computation time.
We adopted a simplified approach in which stellar evolution recipes are applied 
to the initial set up, passively evolving the GC stellar population until 
an age of 12 Gyr. Then, the Monte Carlo simulation is run with these updated
initial conditions and without stellar evolution for an arbitrary long time.
We halt our simulations after 10.5 initial half-mass relaxation times which
corresponds to the dissolution time of most of our simulations.
In this way, each snapshot of the simulation can be considered as the outcome of
a 12 Gyr-long evolution of a GC in a different dynamical stage.
Of course, this is a rough approximation because the effect of stellar
evolution is diluted over the entire life of a real GC, slowly changing the
extent of the mass spectrum and the evolution of single and binary stars.
However, this approximation 
does not introduce any bias in the BSS radial distribution. Indeed, each BSS and its 
progenitor binary system have the same total mass and are therefore subject to the same 
dynamical evolution. So, since binaries are converted into BSS with the same efficiecny at 
all radii, the timing of BSS formation is irrelevant for our purpose. Moreover, stellar evolution driven mass-loss mainly occurs in the first Gyr of evolution
and it is therefore not expected to be significant on the long-term evolution of
the simulation \citep[but see][]{2015MNRAS.449L.100C}. 

Stars were extracted from an Initial Mass Function (IMF) defined between 
0.08 and 120 $M_{\odot}$ and distributed in phase-space following a
\citet{1966AJ.....71...64K} distribution function. No primordial mass segregation is included.
A population of primordial binaries has been
also simulated by random pairing stars from the IMF and assuming the period and
eccentricity distribution of \citet{1991A&A...248..485D}. From the initial
library of binaries we retained only those whose corresponding semi-major axis
is comprised between the Roche-lobe overflow separation \citep{1988ApJ...334..688L} and
the hard-soft boundary ($a_{max}=Gm_{1}m_{2}/2\langle m \rangle \langle
\sigma^{2} \rangle$; where $m_{1}$ and $m_{2}$ are the two components of the
binary and $\langle \sigma^{2} \rangle$ is the square of the 3D velocity dispersion
averaged over the entire cluster).
The stellar evolution recipes of \citet{2009A&A...507.1409K} have been applied to both
single and binary stars. We assume a 100\% retention fraction of white dwarfs
and different retention efficiencies for NSs and BHs.
The initial number of particles in the {\it main sample} is set to $N=29000$ in all 
simulations. The corresponding total mass of the system, after passive stellar evolution,
turns out to be $M\sim~10^{4} M_{\odot}$.
The simulated GC moves on a circular orbit within a tidal field generated by a point-mass of
$M_{G}=10^{10} M_{\odot}$. 

We run a set of simulations starting from different initial conditions. 
For all simulations we adopted a half-mass radius $r_{h}$=6 pc and assumed a
100\% retention fraction for white dwarfs (resulting in an initial mass fraction of
$\mu_{WD}\equiv M_{WD}/M=24\%$).
We defined a reference simulation in which particles are extracted from a \citet{2002Sci...295...82K}
IMF, distributed according to a \citet{1966AJ.....71...64K} model with a central adimensional
potential $W_{0}=5$, a fraction of binaries $f_{b}=5\%$, no heavy remnants
($\mu_{NS,BH}=0$), and move at a distance of
$R_{G}=5~kpc$ from the point-mass galaxy. This implies a tidally filling
($r_{t}\sim r_{J}$) cluster
at the beginning of the simulation. Then we set the other simulations
by changing one of the above parameter at time.  
The entire set of simulations is summarized in Table 1.
To define the core-collapse time we adopted the technique described in \citet{2007MNRAS.374..344T} 
who place this event at the time when a break in the
core-concentration evolution become noticeable. This definition, while
non-optimal in case of an extreme density contrast (see below), has been found to be
adequate in most of our simulations, being able to define this stage of 
evolution even in simulations with a significant fraction of primordial binaries 
and immersed in a strong tidal field which do not show a sharp transition in the
core size evolution (see Sect. \ref{cons_sec}).

In Fig. \ref{all} the evolution of the mass, the
core concentration (defined as $log(r_{h}/r_{c})$), the mass function slope
$\alpha$, the Roche-Lobe filling
factor (defined as $log(r_{h}/r_{J})$), the fraction of remnants and binaries is shown for the entire set of
simulations as a function of the initial half-mass relaxation time.
In the above definitions, we adopted the core radius ($r_{c}$) definition by 
\citet{1985ApJ...298...80C} and $r_{J}$ is the Jacobi radius \citep{1962AJ.....67..471K}.
It can be seen that the various parameters evolve differently according to the
initial conditions. For instance, the core concentration varies by an order of magnitude among the
whole set of simulations. On
the other hand, simulation w5rg10, at odds with all the other
simulations, underfills its Roch-lobe during its entire evolution thus retaining
most of its mass and maintaining almost constant its global mass function. 
Instead, the heavy remnants present in simulation w5rg5bh 
quickly sink into the cluster core and act as an energy reservoir which 
accelerates the response of the core and lead to a general expansion of the
system. Note that, because of the core radius definition adopted here
\citep[from][]{1985ApJ...298...80C}, 
in this last simulation the core is almost coincident with the region containing
the entire BH population. This leads to the occurrence of core-collapse at early
epochs, while visible stars maintain a more extended distribution.

We show in Fig. \ref{time} the comparison of the three dynamical timescales
($t/t_{rh}(t)$, $S$ and $t/t_{cc}$) in all our simulations. It can be seen that,
as discussed in Sect. \ref{cons_sec}, there is not a universal correspondence
between these three indicators when different initial conditions are considered.
In particular, the core-collapse occurs significantly at smaller values of both
$t/t_{rh}(t)$ and $S$ in simulation w5rg5bh (because of the early response of 
the core due to the kinetic energy injected by heavy remnants, see above) and 
w5rg10 (because of its longer relaxation time) than in the other ones.
In the same way, the parameter $S$ grows faster in simulation w5rg10 than in all
the others subject to a strong tidal field. This occurs because
simulation w5rg10 evolves almost in isolation and balance its slow mass-loss
with an expansion keeping the half-mass relaxation time almost constant, while the 
other simulations are tidally limited (i.e. evolve at constant total density)
and their half-mass relaxation time decrease with their mass. So, in the advanced 
stages of evolution of these
simulations the instantaneous half-mass relaxation times is smaller than the
its average value.

\section{BSS radial distribution}
\label{rad_sec}
 
\begin{figure*}
 \includegraphics[width=\textwidth]{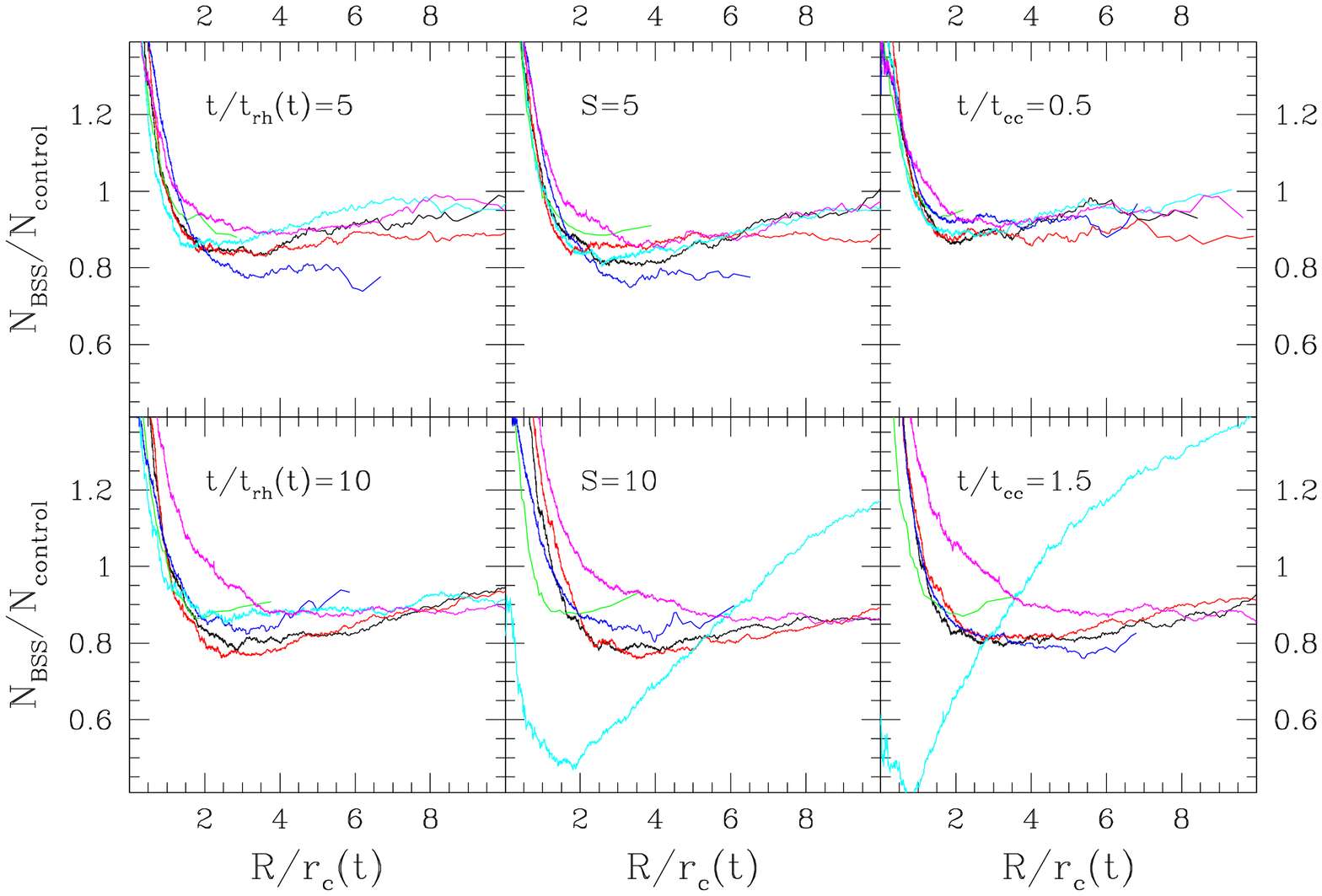}
 \caption{Comparison of the normalized ratio of {\it BSS} and {\it control} 
 sample particles in our set of simulations. The distributions measured at two  plotted in
 different values of $t/t_{rh}(t)$ (left panels), $S$ (middle panels) and
 $t/t_{cc}$ (right panels) are shown. The adopted colour code is the same of Fig. \ref{all}.}
\label{rap2}
\end{figure*}

\begin{figure}
 \includegraphics[width=8.6cm]{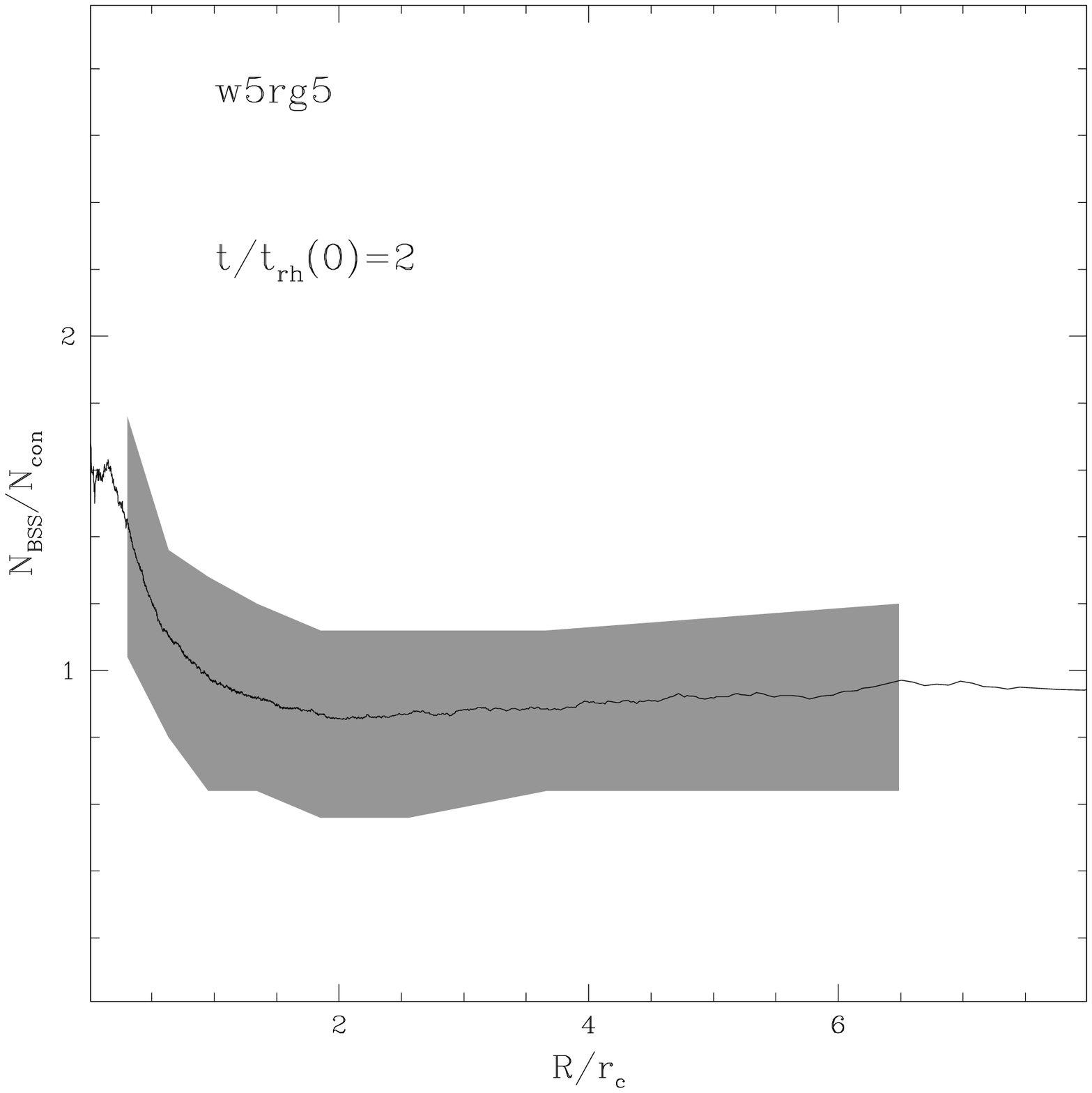}
 \caption{Relative fraction of BSS as a function of the projected distance for
 the snapshot at $t/t_{rh}(0)=2$ of the simulation w5rg5 (solid line). The grey area
 indicate the region including 68\% of the measures preformed in $10^{4}$ 
 random realizations.}
\label{monte}
\end{figure}


As a first analysis, we investigate the behaviour of the relative distribution
of {\it BSS} and {\it control} sample particles in our simulations at different
epochs, and compare it with what observed in real GCs.

In principle, we could use the positions of the {\it BSS} and {\it control}
sample particles to determine the radial distribution of these two populations
in each time-step.
As explained in Sect. \ref{met_sec}, the position of a particle in a Monte
Carlo simulation is chosen randomly along its orbit with a probability
proportional to the time spent in that position. This involves a stochasticity
which adds a small, but significant, noise in the derived radial distributions. 
To eliminate this source of uncertainty, we derived the radial distributions of
{\it BSS} and {\it control} sample particles directly from their energies and
angular momenta by summing the normalized probabilities of each particle to be
found at a given distance
$$N(r)=\displaystyle\sum_{i=1}^{N}\frac{T_{i}(r)}{\int_{0}^{r_{t}} T_{i}(r) dr}$$
where $T_{i}(r)$ is the time spent by the i-th particle at the distance $r$, and
can be calculated using eq. 4 of \citet{2014MNRAS.443.3513S}.
The 3D density of {\it BSS} and {\it control} sample particles have been then
projected into the plane of the sky and their ratio has been calculated. For a
homogeneous comparison with observations, the above ratio has been calculated by
normalizing the above derived densities to the number of {\it BSS} and 
{\it control} sample particles

$$\frac{N_{BSS}}{N_{con}}(R)=\frac{N_{BSS}(R)~N_{con}^{tot}}{N_{con}(R)~N_{BSS}^{tot}}$$

The normalized $N_{BSS}/N_{con}$ ratio calculated in different stages of
evolution is shown in Fig. \ref{rap} for simulation w5rg5 as a function of the projected distance 
from the cluster center in units of both the core and the half-mass radius.
The relative number of BSS shows a clearly defined behaviour and evolution:
while at the beginning of the evolution it is almost flat (reflecting the
absence of primordial mass segregation), a central peak develops whose intensity
grows with time. This is the natural outcome of dynamical evolution since BSS,
which are more massive than {\it control} sample stars, segregate into the
cluster core in a timescale comparable to the core relaxation time. 
 At larger
distances the relative fraction of BSS has a minimum at $N_{BSS}/N_{con}\sim0.8$
and then slowly increases at increasing distances reaching again the primordial
value close to the tidal radius. The BSS distribution remains bimodal during
the following evolution being therefore a permanent feature, at odds with what
found by \citet{2017MNRAS.471.2537H}.

To understand the physical reason of such a behaviour we show in Fig. \ref{ene} 
the distribution of orbital energies (which together with the angular momenta
determine the radial distribution of particles) of {\it BSS} and {\it control}
sample particles in different stages of evolution. Note that while at the
beginning the two populations are distributed in a similar way in the energy
domain, as evolution proceeds BSS progressively migrate toward low energies
(i.e. inner orbits) producing the central peak in the relative fraction.
However, a significant fraction of BSS conserve their original energy creating a
bimodal distribution which reflects into the radial distribution. These BSS
indeed spend most of their life in the peripheral region of the cluster, where
the local relaxation time is longer than the cluster age, without interacting
significantly with other cluster stars. This process is also enhanced by the
presence of a mass spread in the {\it BSS} sample: the dynamical friction
timescale is indeed inversely proportional to mass and massive BSS tend to sink
into the cluster core faster than low mass ones. 
The same bimodal distribution is not apparent in the {\it control} sample particles 
because {\it i)} their mass constrast with respect to the other clusters stars 
is smaller than that of BSS, and {\it ii)} they do not contain a spread of masses.
 
Fig \ref{rap} shows that, although the minimum of the distribution seems to be 
significantly less pronounced in the models considered here than that detected in the observations, its position, when expressed in units of core radii ($R_{min}/r_{c}$), 
appears to progressively migrate 
outward. This migration is not evident in the right panel when the radial 
position of the minimum is expressed in unit of the half mass radius
$R_{min}/r_{h}$. This is visible in Fig. \ref{rad} 
where the evolution of $R_{min}$ is shown for simulation w5rg5. 
It is apparent that the behaviour of the minimum is quite noisy and it is 
difficult to evaluate its decreasing rate at increasing time. 
In fact the width of the region where the minimum is located increases 
with time making its exact localization difficult \citep{2015ApJ...799...44M}. 
As reference the evolution of both 
 $r_{c}$ and $r_{h}$ are also shown. From a qualitative comparison, the 
 decreasing trend of the minimum appears to be more similar to that followed by 
 the
 half-mass radius than to that of the core radius. This suggests that a 
 negligible radial trend as a function of time should be expected when the 
 $R_{BSS}$ distribution is plotted as a function of $R_{min}/r_{h}$, at odds 
 with what found in the observations (see Fig. \ref{rmin}) that show 
 that the radial migration of the minimum is independent of the normalisation 
 of the radial coordinate. 
However it is important to note that $t/t_{rh,0}$ is not known for the 
observed clusters in Fig. \ref{rmin}, nor is our suite of models aimed at 
reproducing their individual properties. Hence further study is required to 
determine if the behaviour of $R_{min}/r_{h}$ in our simulations is in 
disagreement with observations or just an artifact of the limited sample of 
observed clusters and/or the considered range of initial conditions in our simulations.
It should be also noted that, at odds with simulations, observers usually estimate 
the core and half-mass radius through a fit of the surface brightness or star 
counts profile of the brightest population 
with some analytic model. This can introduce a time-dependent bias in the 
determination of these scale radii. To check the impact of such an uncertainty, 
we determined $r_{c}$ and $r_{h}$ in all the snapshots of simulation w5rg5 
by fitting their projected density profile with a set of \citet{1966AJ.....71...64K} models. 
For this purpose, in each snapshot the 3D positions of unevolved stars with 
mass $M>0.7~M_{\odot}$ have been projected along one direction and their cumulative 
radial distribution have been compared with that of \citet{1966AJ.....71...64K} models 
with different concentrations. The best fit model has been chosen as the one 
providing the best KS statistic. 
The $r_{c}$ and $r_{h}$ of the best fit models 
are overplotted to the actual ones in Fig. \ref{rad} as red lines. 
It is apparent that
both $r_{c}$ and $r_{h}$ estimated through the fit of bright 
stars decrease faster than their actual values until core-collapse. 
However, also in this case, the decreasing trend of $R_{min}$ and $r_{h}$ still 
appears to be similar, at odds with 
what shown by the observations in Fig. \ref{rmin}. Thus this issue remains open.

In Fig. \ref{rap2} we compare the relative fraction of {\it BSS} and {\it
control} sample particles as a function of distance predicted by different
simulations. Note that, although the height of the central peak varies in
different simulations, the dip in the radial distribution is apparent in all of
them but for the case of w5rg5bh. The heavy remnants present in this last
simulation indeed dominate the core and interact with the BSS sunk into the
core, thus reducing the effect of mass segregation. Note also that for
simulation w3rg5 the bimodal distribution appears only after core-collapse.

%
The bimodal behaviour of the relative fraction of BSS is qualitatively similar 
to that observed in real GCs by \citet{2012Natur.492..393F}. However, the minimim observed in many GCs reaches sometimes very
low values (down to $N_{BSS}/N_{con}\sim0.1$; see their Fig. 2) which are never
reached in our simulations. 

In order to investigate the effect of Poisson fluctuations we calculated the radial 
profile of the relative fraction of BSS also from
$10^{4}$ random extractions of 100 BSS and 500 control particles. 
The result for the snapshot at $t/t_{rh}(0)=2$ of simulation w5rg5 is shown in 
Figure \ref{monte}, where we highlighted in grey the region where 68\% of the 
extractions are located. We see that while no significant bias is apparent at 
any radius, the uncertanties in the population ratios can be significantly 
larger that those discussed above. A similar amount of noise is expected also in the simulations by
\citet{2017MNRAS.471.2537H} which contain samples of BSS and control stars of similar
sizes. These fluctuations allow the detection of the growing trend of the
relative fraction of BSS in only a fraction of the analysed snapshots,
erroneously leading to the interpretation of this feature as transient. 
We also notice that the identification of the minimum in such condition is even 
more difficult, possibly producing false detection.

It is worth noting that in our simulations we considered only BSS formed through 
the evolution of primordial binaries. In principle, a contribution to the BSS 
population could be also given by the collisions between single stars and by 
binaries. However, "collisional" BSS are expected to form 
in the cluster center and remain confined within the core during their entire evolution. 
So, this effect would increase the height of the central peak in the $N_{BSS}/N_{control}$ ratio 
without affecting its radial trend outside the core.

\section{BSS segregation indicators as "dynamical clocks"}
\label{coef_sec}

\begin{table}
 \centering
 \label{tab:table2}
  \caption{horizontal spread of the three indicators of dynamical evolution in Fig.s \ref{coeff} and \ref{bianch}.}
  \begin{tabular}{@{}lcccr@{}}
  \hline
                 & value & $\sigma(t/t_{rh}(t))$ & $\sigma(S)$ & $\sigma(t/t_{cc})$\\
 \hline
 $R_{min}/r_{c}$ &   3   & 3.11 (7.7)        & 2.02 (7.0) & 0.27 (1.7)\\
 $A^{+}$         &  0.07 & 1.88 (5.4)        & 1.38 (4.0) & 0.21 (0.5)\\
 $c_{k}$         &  0.5  & 0.67 (2.0)        & 0.48 (1.4) & 0.35 (0.9)\\
\hline
\end{tabular}
\end{table}

\begin{figure*}
 \includegraphics[width=\textwidth]{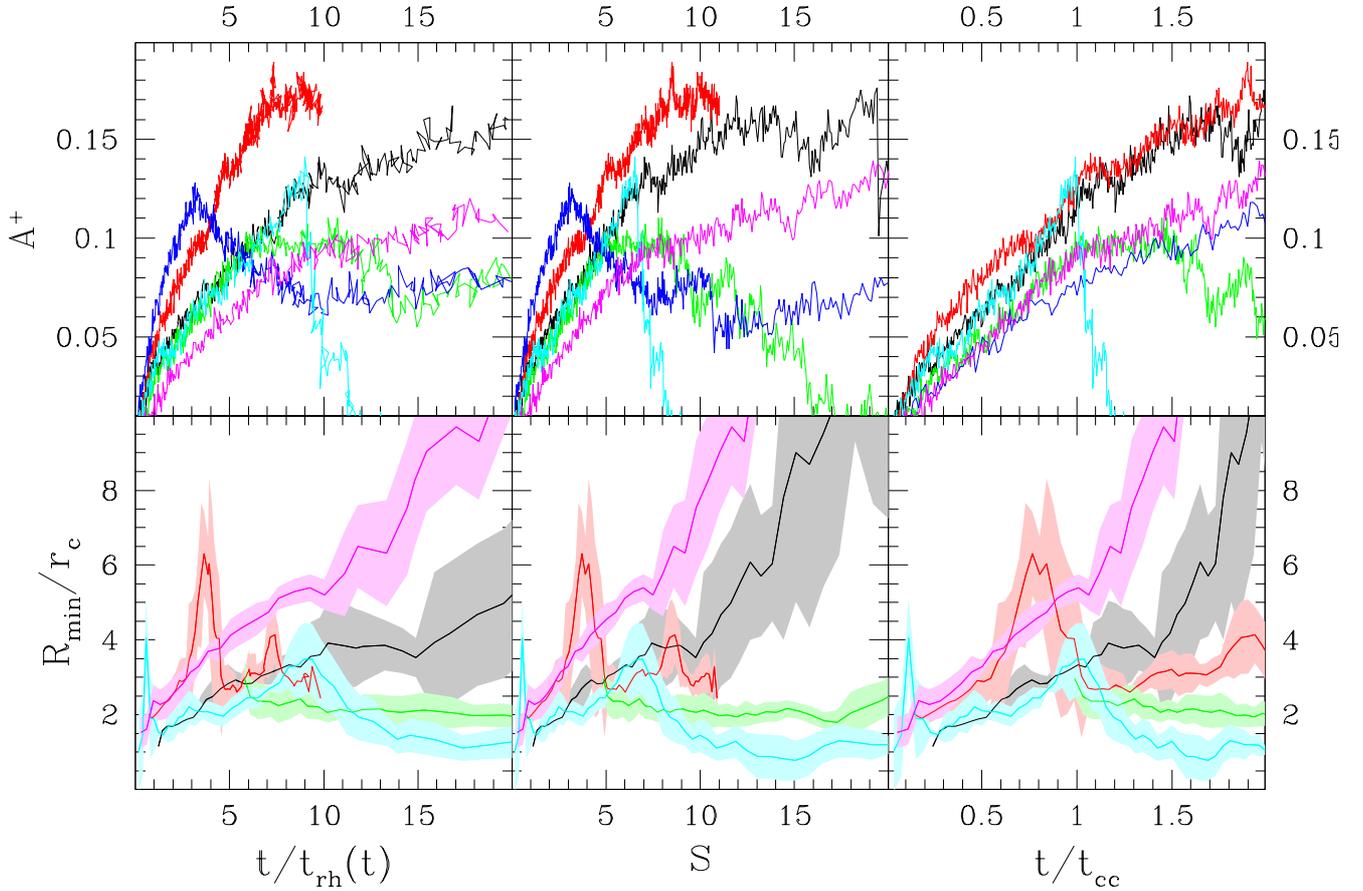}
 \caption{Evolution of the $R_{min}/r_{c}$ (bottom panels) and $A^{+}$ (top
 panels) parameters as a function
 of the three dynamical evolution time-scales defined in Sect. \ref{cons_sec}. $t/t_{rh}(t)$ (left panels),
 $S$ (middle panels) and $t/t_{cc}$ (right panels). The adopted colour code is
 the same of Fig. \ref{all}. In the bottom panels the 1$\sigma$ spread around
 the mean trend is marked by the shaded area. The 
 simulation w5rg5bh (blue line), for which a unimodal distribution is found during 
 its entire evolution, is not present in the bottom panels.}
\label{coeff}
\end{figure*}

\begin{figure*}
 \includegraphics[width=\textwidth]{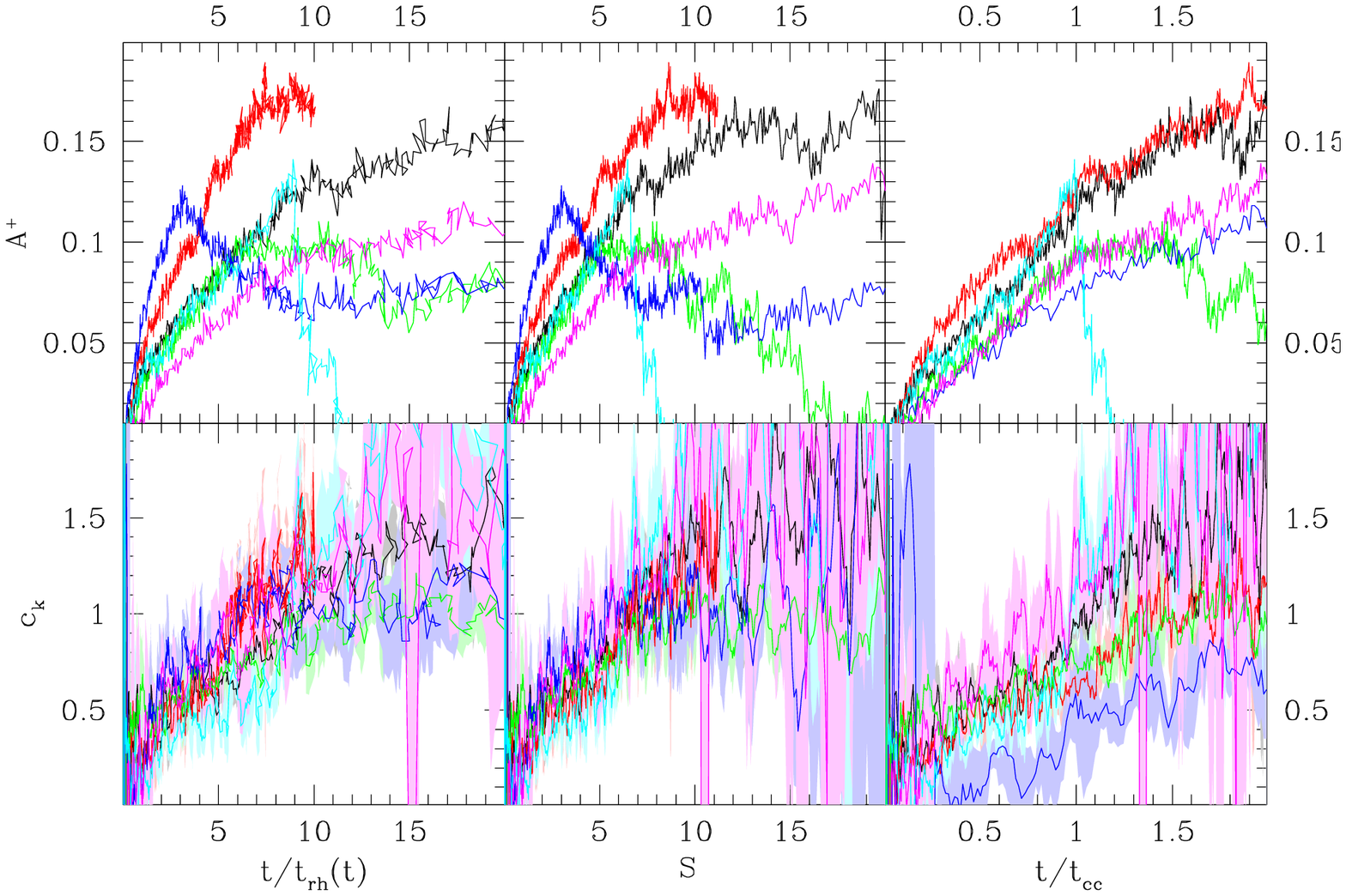}
 \caption{Evolution of the $c_{k}$ (bottom panels) and $A^{+}$ (top
 panels) parameters as a function
 of the three defined dynamical time-scales $t/t_{rh}(t)$ (left panels),
 $S$ (middle panels) and $t/t_{cc}$ (right panels). The adopted colour code is the same of Fig. \ref{all}. 
 In the bottom panels the 1$\sigma$ spread around
 the mean trend is marked by the shaded area.}
\label{bianch}
\end{figure*}

Once clarified the nature of the dip in the relative fraction of BSS we
investigated the behaviour of this feature, as well as the parameter $A^{+}$,
during the dynamical evolution of our simulations.

In principle, an ideal empirical "dynamical age" indicator should {\it i)} show a monotonic behaviour in terms of (at least) one the three time-scales 
defined in Section 2, and {\it ii)} show a good level of independence on initial conditions.
According to its definition \citep{2016ApJ...833..252A} and its recent applications \citep{2016ApJ...833L..29L, 2017ApJ...839...64R,2018ApJ...860...36F},
the parameter $A^{+}$ has been calculated as the area delimited by the 
normalized cumulative radial 
distributions of {\it BSS} and {\it control} particles, when the projected radial 
distances are expressed in logarithmic units.
In Fig. \ref{coeff} the evolution of the two considered parameters as a function
of the three considered dynamical evolution time-scales is shown for all our
simulations. The behaviour of $R_{min}/r_{c}$ turns out to be quite noisy, even in our simulations where the
number of {\it BSS} and {\it control} sample particles are oversampled by about
2 orders of magnitudes. Thus, in order to better delineate the overall trend of this parameter as a function of different dynamical tima-scales,
 the best fitting curves to the simulations and their $1\sigma$ spreads are plotted in the Figure. As apparent, the parameter shows
a generally increasing profile at early epoch with a decrease occurring in some simulations
after core-collapse. 
Moreover, as shown by \citet{2012Natur.492..393F}, highly dynamically evolved 
clusters show an unimodal BSS distribution, hence no minimum is expected to be 
detected in those clusters. This is also true for dynamically unevolved 
clusters, where the normalized BSS distribution is flat and again no minimum 
can be detected. All these limitations, combined with the intrinsic difficulties in 
detecting the minimum, makes the use of 
$R_{min}/r_{c}$ as an indicator of dynamical age not optimal.

On the other hand, the $A^{+}$ parameter shows a similar behaviour but with much more 
clear and well-defined trends. Indeed, this
 parameter monotonically grows with time until core-collapse, while after this stage
the behaviour of $A^{+}$ significantly vary in simulations with different 
initial conditions. In particular, after core-collapse the energy budget of the
core (where most BSS reside) is supported by the interactions involving
binaries. BSS, being more massive than other control sample particle, are more
subject to such interactions and are scattered at large distances. This effect
can produce a decrease of the mass segregation index $A^{+}$ which is more
prominent when more energetic interactions (like those occurring in simulation w5rg5nobin
starting without primordial binaries) occur. 
Note also that, while the 
evolutionary paths of $A^{+}$ predicted by different
simulations have a significant spread when expressed in units of $t/t_{rh}(t)$ 
and $S$, a relatively small spread is apparent in terms of $t/t_{cc}$ for $t<t_{cc}$.
So, although  the criteria defined at the beginning of this section are not strictly 
satisfied, the $A^{+}$ parameter could be used as a proxy for the time needed to reach
core-collapse in those GCs who have still not experienced this event.

For comparison, in Fig. \ref{bianch} the evolution of the $A^{+}$ parameter is compared with that of the 
"kinematic concentration" parameter ($c_{k}$) which has been also proposed as an indicator of dynamical 
evolution by \citet{2018MNRAS.475L..96B}.
It can be seen that the kinematic concentration satisfies both the criteria defined above
along the entire evolution of all our simulations. On the other hand, although observational uncertainties 
have not been simulated here, it is subject to a significant noise and can be 
considered independent on initial condition only when expressed in terms of 
$t/t_{rh}(0)$ and $S$. Beside intrinsic fluctuations, part of the noise 
apparent in Fig. \ref{bianch} is due to the low-number statistics and could be 
significantly reduced in observations of real GC. We note however that, at odds with the indicators based on the radial 
distribution of BSS, the derivation of the $c_{k}$ parameter is observationally challenging because 
it requires the measure of velocities of 
stars in different mass regimes along the Main Sequence \citep[a task which became feasible only 
recently;][]{2014ApJ...797..115B,2018ApJ...853...86B,2016A&A...588A.149K,2018MNRAS.473.5591K}, and 
the derivation of individual masses through the comparison with 
suitable stellar models.

For a quantitative analysis, we calculated the horizontal spread in the
considered set of simulations in Fig.s
\ref{coeff} and \ref{bianch} at specific values of $R_{min}/r_{c}$, $A^{+}$ and
$c_{k}$ in units of $t/t_{rh}(t)$, $S$ and
$t/t_{cc}$. Although our simulations do not span the entire range of initial
conditions covered by real GCs, these spreads can be considered as 
the typical uncertainty in the derived "dynamical age" for each of the
considered indicators. We chose as reference values $R_{min}/r_{c}=3$,
$A^{+}=0.07$ and $c_{k}=0.5$ which roughly corresponds to the values of these
parameters in the reference simulation (w5rg5) at the time $t/t_{rh}(0)=2.2$. 
The measured standard
deviations and the maximum differences among the various simulations are listed in Table 2. It
can be noted that, while indicators based on the radial distribution of BSS
cover a range of several $t/t_{rh}(t)$ and S, the kinematic concentration $c_{k}$ 
has a significantly smaller spread. On the other hand, in terms of $t/t_{cc}$, the situation is reversed with
$A^{+}$ providing the smallest spread. This is somehow expected because of the
different location of the tracers adopted by the above indicators: while the
$c_{k}$ parameter is measured in terms of global quantities (and it is therefore
well correlated with $t/t_{rh}$), those parameters
based on the BSS radial distribution are more sensitive to the dynamical
evolution of the core (and better correlated with $t/t_{cc}$), where most BSS reside.

\section{summary}
\label{summ_sec}

Through a set of Monte Carlo simulations starting with different initial
conditions (strength of the tidal field, concentration, binary and heavy remnant 
fraction and mass function) we found that dynamical evolution can naturally
reproduce a bimodal radial distribution of the BSS population.
In particular, the presence of a dip quickly emerges in most simulation and remains visible along the
entire simulation. 

This feature is produced by the progressive migration of BSS
in the energy domain determining a bimodal orbital energy distribution.  
A similar dip has been already identified in recent N-body and Monte Carlo \citep{2015ApJ...799...44M,2017MNRAS.471.2537H} simulations.
In particular, \citet{2017MNRAS.471.2537H} interpreted the discontinuous detection of this feature in subsequent snapshots of their simulations as an evidence of its transient nature. 
This is the first time that the significance and stability of this feature, is verified with a large set
of Monte Carlo simulations. 
Indeed, while the N-body simulations of \citet{2015ApJ...799...44M} and \citet{2017MNRAS.471.2537H} both contain 3 times 
more particles, they include only a small number of BSS which hamper the possibility to draw any firm 
conclusion on this issue. On the other hand, the artificial oversampling of the distribution function of BSS allowed us to
determine the radial distribution of BSS with thousands of particles tearing down 
the effect of statistical fluctuations. 

However, although such a feature is 
qualitatively similar to those observed in real GCs
\citep[see][]{2012Natur.492..393F}, a few details remain not reproduced in the 
simulations.
First, the migration of the position of the dip seems to be sensitive 
to the normalisation radius ($r_c$ or $r_h$) at odds with observations, where 
the position of the minimum follow the same evolution regardless of
the radial normalization (see Fig. \ref{rmin}).

Second, the depth of the dip is less pronounced than what found in the observations, 
as already found in the N-body and Monte Carlo simulations by \citet{2015ApJ...799...44M} and \citet{2017MNRAS.471.2537H}. 
While our simulations do not account for the effect of stellar evolution,
simulate clusters with a mass which is more than one order of magnitude smaller
than those of real GCs and cover only a small portion of the parameter space of
initial conditions, it seems difficult to reproduce the almost complete lack of
BSS at intermediate distances observed in many GCs \citep[those belonging to the 
"Family II" defined in][]{2012Natur.492..393F}. 
Moreover, as demonstrated in Fig 8, when a realistic number of BSS
and RGB stars is considered, the expected
uncertainties in the measured normalized ratio are $\sim$1.5 times larger than
the signal predicted by our simulations. So, if such a large uncertainties can explain the observed bimodal trend in some 
individual GC, this feature should not be easily observable in a large sample of GCs.
In summary, although our simulations seem to identify the physical process 
leading to a bimodal BSS radial distribution, it is not clear if additional 
phenomena occurs making this feature as strong as observed in real GCs.

The behaviour of the $R_{min}/r_{c}$ and $A^{+}$ parameters 
\citep[recently proposed by][]{2016ApJ...833..252A} has been also investigated 
and compared with that of the "kinematic concentration" parameter 
proposed by \citet{2018MNRAS.475L..96B}. 
The $R_{min}/r_{c}$ appears to be extremely noisy also in our
idealized simulations containing samples of thousands of BSS and control sample
particles. On the other hand, if we restrict our analysis to the
epoch preceding core-collapse, the evolution 
of the $A^{+}$ parameter shows a relatively small spread when expressed in terms
of the core-collapse time ($\sigma(t/t_{cc})\sim0.2$)
regardless of the initial conditions of the simulation.
The same conclusion has been reached by \citet{2016ApJ...833..252A} who however considered only
N-body simulations without binaries and did not follow the post-core-collapse
evolution. 

Thus, although this parameter does not satisfy the general
criteria defined above along the entire evolution, it could 
be considered a quite promising 
indicator of the evolution of a GC along its path toward core-collapse at least 
during its pre-core-collapse phase. 
On the other hand, the $c_{k}$ parameter is found to be more effective in 
tracing the global relaxation of the cluster, while being less sensitive to the 
core evolution. These parameters provide therefore complementary information on 
the cluster overall dynamical evolution.

In the future, it will be interesting to verify whether the same behaviour is also 
seen in the observations. At the moment, the observational facts suggested that
a negligible fraction of BSS is found in the external regions of post-core-collapse GCs 
and that the GC with the largest measured value of $A^{+}$ \citep[in a sample of 
48 GCs;][]{2018ApJ...860...36F} is indeed the post-core-collapse cluster NGC6397.
In comparison with other indicators of dynamical evolution 
\citep[defined as a function of global quantities and therefore sensitive to 
the general efficency of two-body relaxation;][]{2018MNRAS.475L..96B,2017MNRAS.464.1977W} 
the $A^{+}$ parameter is more sensitive to the process of relaxation 
occurring in the core where most BSS reside. All these indicators can be 
therefore used in a complementary way to study different aspects of the dynamical 
evolution of GCs.

It will be also interesting to investigate if differences in initial condition of the binary population possibly
present at the formation of the GC stellar populations 
could possibly affect the long-term evolution of its BSS population.   
Indeed, any difference in initial condition of the binary population possibly
present at the formation of the GC stellar populations 
could reflect on the long-term evolution of its BSS population. In this context,
different binary fractions have been observed in Na-rich and Na-poor stars in
several GCs \citep{2015A&A...584A..52L} whose evolution could lead to bimodal trends
in their radial relative fraction \citep{2015MNRAS.449..629H,2016MNRAS.457.4507H}. 
However, the entire process of star formation in GCs  
occurs at early epochs ($>10^10$ yr) and in a short time-scale \citep[$<10^{8}$ yr]{2008MNRAS.391..354R,2008MNRAS.391..825D}, and 
it is therefore difficult that initial conditions have not been erased by 
two-body relaxation. In this respect, it would be 
interesting to investigate the evolution of the BSS population in GCs 
characterized by multiple episodes of early star formation.
From the observational side, any possible link between the radial distribution 
of BSS and their chemical properties would be helpful.

\section*{Acknowledgments}
We warmly thank Michele Bellazzini, Enrico Vesperini and Anna Lisa Varri for 
useful discussions. We also thank the anonymous referee for his/her helpful comments and suggestions.

\label{lastpage}

\end{document}